**Mechanisms of weak thickness dependence of the critical current density in strong pinning *ex situ* metal-organic-deposition route YBa$_2$Cu$_3$O$_{7-x}$ coated conductors**

S. I. Kim,[1] A. Gurevich[1], X. Song,[1] X. Li,[2] W. Zhang,[2] T. Kodenkandath,[2] M.W. Rupich,[2] T. G. Holesinger,[3] and D. C. Larbalestier[1]

[1]Applied Superconductivity Center, University of Wisconsin, Madison, WI 53706

[2]American Superconductor Corporation, Westborough, MA 01581

[3]Los Alamos National Laboratory, Los Alamos, NM 87545

ABSTRACT

We report on the thickness dependence of the superconducting characteristics including critical current $I_c$, critical current density $J_c$, transition temperature $T_c$, irreversibility field $H_{irr}$, bulk pinning force plot $F_p(H)$, and the normal state resistivity curve $\rho(T)$ measured after successive ion milling of ~ 1 μm thick high $I_c$ YBa$_2$Cu$_3$O$_{7-x}$ films made by an *ex situ* metal-organic deposition process on Ni-W rolling-assisted biaxially textured substrates (RABiTS$^{TM}$). Contrary to many recent data, mostly on *in situ* pulsed laser deposition (PLD) films, which show strong depression of $J_c$ with increasing film thickness $t$, our films exhibit only a weak dependence of $J_c$ on $t$. The two better textured samples had full cross-section average $J_{c,avg}$(77K,0T) ~ 4 MA/cm$^2$ near the buffer layer interface and ~3 MA/cm$^2$ at full thickness, despite significant current blocking due to ~30% porosity in the film. Taking account of the thickness dependence of the porosity, we estimate that the local, vortex-pinning current density is essentially independent of thickness, while accounting for the additional current-blocking effects of grain boundaries leads to local, vortex-pinning $J_c$ values well above 5 MA/cm$^2$. Such high local J$_c$ values are produced by strong three-dimensional vortex pinning which subdivides vortex lines into weakly coupled segments much shorter than the film thickness.

# I. INTRODUCTION

Second generation high temperature superconducting (HTS) wires based on coated conductors (CC) can meet the demand for high critical currents $I_c$ for high power applications in motors, generators, transformers, fault current limiters, superconducting magnetic energy storage, and power transmission lines [1]. An obvious route to high $I_c$ is to increase the thickness $t$ of the $YBa_2Cu_3O_{7-x}$ (YBCO) layer, but significant reduction of the critical current density $J_c$ with increasing $t$ has often been observed [2-9]. Whether this reduction is due to fundamental vortex physics or thickness-dependent material microstructure is not yet fully clear. For instance, portions of the YBCO layer with reduced local $J_c$ used to be rather common, both at the surface and at the buffer layer interface [3], although they are no longer pronounced in high-performance CCs. Advances in *ex situ* continuous methods of YBCO growth are raising questions about the effect of the growth process on transport properties. Indeed, it has been tacitly assumed until recently that the particular method of YBCO growth, if properly done, does not significantly affect the superconducting properties, and therefore that results for the widely studied *in situ* pulsed laser deposition (PLD) films should be generally applicable to the many newer types of YBCO films. But there is increasing evidence that *ex situ* and *in situ* methods are not fully equivalent [10-12]. In this much more diverse era of industrially produced high $J_c$ coated conductors there is thus an important need to much better understand the thickness dependent properties of *ex situ* films.

Because typical YBCO layers in practical CCs are 1-2 μm thick, the route to high $I_c$ involves not only maximization of the $J_c$ but also minimizing the degradation of $J_c$ with increasing thickness. Such degradation of $J_c(t)$ of YBCO films has been well documented in the literature. For example, the thickness dependent $J_c(t)$, of *in situ* PLD films with $t$ ranging from ~100-500 nm was investigated by van der Beek *et al.* [13] who observed a maximum in $J_c(t)$ at t ~ 100-200 nm. The temperature and field dependences of $J_c(T,B)$ for these films and their high self-field values $J_c(77K)$ ~ 4 MA/cm$^2$ are consistent with a single-vortex pinning by strong pins such as $Y_2O_3$ precipitates [14]. The decline of $J_c$ with increasing $t$ for many PLD films [2-9] usually exhibits an inverse square root like dependence $J_c(t) \propto t^{-1/2}$ which levels off above a critical thickness $t_c$ ~ 1 μm [2-9]. Such a thickness dependence is suggestive of the transition from the 2 dimensional (2D) pinning of rigid vortex lines in films thinner than the longitudinal pinning correlation length $l_c$ to the 3-dimensional (3D) pinning of deformable vortices at $t > t_c$ [14]. It was recently pointed out [15] that $t_c$ can indeed approach a few μm if the collective pinning model incorporates a multi-scale pinning potential appropriate for the strong-pinning second phase precipitates, pores, and correlated defects found in CCs. Strong-pinning defects have a pin interaction range, $r_p$ much greater than the coherence length $\xi$ and produce large plastic deformations of vortices, rather than the small elastic deformations produced by weak, point pins. This strong-pinning model predicts a crossover thickness $t_c$ as large as 1-2 μm, in agreement with the observed $J_c(t)$ dependence of many PLD films [2-8] and qualitatively consistent with many recent studies of the angular dependence of $J_c$ in CCs, which also reveal much evidence for correlated pinning along the *c*-axis in PLD films [16]. This multiscale pinning model also predicts the $t^{-1/2}$ thickness dependence of $J_c(t)$, but the magnitudes of $J_c$ and $t_c$ can be very dependent on the specific pinning microstructure and thus on the film growth process.

Another interpretation of the $t^{-1/2}$ thickness dependence of $J_c(t)$ in PLD films has recently been advanced by Foltyn *et al.* [17]. They attribute this to a local $J_c(z)$ profile (where *z* is distance from substrate) across the film, such that $J_c$ is ~ 7.2 MA/cm$^2$ at the substrate and then linearly decreases out to a thickness of ~0.65 μm, followed by a thickness-independent $J_c$ of ~ 1.4 MA/cm$^2$ at larger thickness. The very high $J_c$ at the bottom of the YBCO layer is attributed to a network of misfit dislocations localized near the interface of the YBCO and the substrate cap layer. Whatever the true cause of this variation, the striking proposal is that the thickness dependence of $J_c$ is due to an important thickness-dependent microstructure.

The above discussion of $J_c(t)$ in various types of YBCO films and CCs raises a question as to whether the transport data set derived from well made PLD films can be generally compared to films grown by other techniques. A strong scientific advantage of PLD films grown on single crystal substrates is that they are normally dense without grain boundaries, so that $J_c = I_c/A$ is generally a good measure of the local flux pinning, where *A* is the whole geometrical cross-section of the superconducting film By contrast, continuous film growth processes made on CC templates often produce films with current-blocking defects, such as pores, cracks, grain boundaries, and a worsening epitaxy as the film thickness increases. For example, all CCs grown on RABiTS™ have significant self field limitation of $J_c$ due to current-blocking grain boundaries [10,18]. In this case the observed $I_c$ is determined not only by local flux pinning in the grains, but also by the effective current-carrying cross-section $A_{eff}$ between higher angle grain boundaries, which may be markedly less than *A*.

The mixture of vortex physics and materials microstructure produced by any particular growth technique is complex and still poorly understood. For example, liquid is now recognized to be present in many growth modes, even *in situ* PLD or electron-beam processes [19], as well as *ex situ* methods involving fluorine-based precursors [19-21]. We observed a strong effect of liquid in an earlier thickness dependent study of $J_c$ in 2 – 3 μm thick films grown by the electron-beam BaF$_2$ deposition route [22] when the *ex situ* conversion conditions produce significant liquid phase [11]. Study of $J_c(t)$ by milling the films to progressively smaller *t* showed depressed $J_c$ values (~0.8-1.2 MA/cm$^2$) which were also rather independent of *t*. Electron microscopy showed a very inhomogeneous microstructure that changed markedly about half way through the film thickness, permitting the independence of $J_c$ on *t* to be ascribed to a bimodal, non-uniform distribution of pinning sites that happened to produce a higher $J_c$ in the upper layer [22]. Microstructural effects are also apparent in another recent study by Emergo *et al.* [8] of up-to-3-μm-thick PLD-deposited YBCO on various miscut SrTiO$_3$ substrates, where different YBCO defect microstructures occur according to the degree of miscut. Although all films showed $J_c$ falling off with increasing $t$ approximately as $\sim t^{-1/2}$, there was both a diversity of $J_c(t)$ behavior and a wide range of $J_c$ which varied from 0.6-1.4 MA/cm$^2$. Adding to this diversity of behavior is a recent thickness-dependent study in our laboratory of ~300 nm *in situ* PLD-grown films grown on single crystal substrates. The $J_c(t)$ behavior is remarkably similar to the data set presented by van der Beek *et al.* [13], but in our case [23] we found evidence that the initial rise in $J_c(t)$ at small $t$ is due to the suppression of thermal fluctuations, as t increases. In summary, given the increasing variety of experimental evidence for strongly varying $J_c(t)$ behavior and materials-related effects, a much better understanding of the relative contributions of physics- and materials-related effects in different types of YBCO films is needed for optimization of the current-carrying capability of coated conductors.

The present work provides a detailed study of $J_c(t)$ of a promising type of metal-organic deposition (MOD) conductors based on *ex-situ* conversion of fluorine-based precursors, which is widely recognized as a cost-effective, high-rate and scaleable technique to produce YBCO CCs [24-26]. High $J_c$ and $I_c$ comparable to, or even better than, those of *in situ* PLD films have been routinely achieved by this method [27,28]. Moreover, variable thickness (0.3-0.9 μm) YBCO CCs processed by the MOD route show fairly constant $J_c$ as a function of thickness [29], making the process very attractive for developing high $I_c$ films. Angular-dependent pinning studies [16,30] and correlations to the microstructure [11,31] support the conclusion that the MOD route produces strong pinning films. A study by Kim *et al.* [10] showed that grain-boundary, current-blocking effects in narrow-tracks of MOD-CCs are much weaker than expected from single grain boundary studies using in *situ* methods. This conclusion is consistent with recent results by Feldmann *et al.* [12] which show that the meandering grain boundaries found in these *ex situ* films block current less than the straighter grain boundaries of *in situ* grown PLD-RABiTS films. In general, it appears that there are fundamental differences between *in situ* and *ex situ* microstructures that significantly affect vortex pinning, grain boundary current limiting effects and thus the thickness dependence of $J_c$ in different types of YBCO films.

In this work we report depth profiling experiments that reveal the thickness dependence of $I_c$, $J_c$, $T_c$, the room temperature resistivity $\rho_{300K}$, resistivity curve $\rho(T)$, the bulk flux pinning force curve $F_p(H)$, and the microstructures of three *ex situ* MOD YBCO films made on Ni-W rolling-assisted biaxially textured substrates (RABiTS[TM]). The significant results of our study are that the average $J_{c,avg}$ remains higher than 3 MA/cm$^2$ after successive ion millings down to the YBCO-substrate interface, that $J_c$, whether averaged over the whole thickness or determined incrementally, shows only a weak thickness dependence, and almost no change of the shape of $F_p(H)$ with thickness is seen. This small variation of through-thickness $J_c$ agrees with prior *variable-thickness-sample* studies on the same type of film [29], which showed only a weak $J_c$ reduction with increasing *t*. However, we also observe a significant through-thickness variation of room temperature resistivity $\rho_{300K}$ and pore density. If the through-thickness variation of current-carrying cross-section caused by varying porosity is taken into account, the local incremental $J_c$ shows even weaker thickness dependence. High $J_c$ values observed in this work require strong flux pinning which may arise from the pores, insulating precipitates, and planar intergrowths found in our films. The recent works of Civale et al. [30] and Holesinger et al. [31] correlate a strong peak in the angular dependence of $J_c$(H) for H parallel to the ab planes to a high density of *ab*-plane stacking faults. Our results, which strongly differ from data taken on many PLD YBCO films, suggest that fine details of the microstructure of different types of YBCO film play a large role in determining the flux pinning behavior and thickness dependence of $J_c$. For MOD films it appears that inherently strong vortex pinning leads to an essentially thickness-independent flux pinning $J_c$ that is partially degraded by current-blocking porosity that builds up in the thicker portions of the film. The deconvolution of the effective area carrying current, due to a thickness dependence of this porosity and also to a field- and temperature-dependent effective cross-section produced by the partial blocking effects of low angle grain boundaries makes estimates of the true flux pinning current density magnitude inherently uncertain, a point that we return to in the later discussion.

## II. EXPERIMENTAL DETAILS

Three different MOD YBCO samples were studied. The total measured thicknesses ranged from 1.0-1.3 μm. Buffer layers of $CeO_2$/YSZ/$Y_2O_3$/Ni were first deposited on Ni5at.%W RABiTS$^{TM}$, then YBCO layers were formed by MOD using a trifluoroacetate (TFA)-based precursor. The precursor was decomposed to a nominal mixture of $BaF_2$, CuO and $Y_2O_3$, which was then continuously converted to the superconducting YBCO phase in a humid, low oxygen partial pressure environment at 700 – 800°C, as described elsewhere [26,29].

YBCO bridges ~300 μm wide and ~500 μm long were cut by Nd-YAG (yttrium aluminum garnet) laser ablation so as to restrict $I_c$ to < 10 A at full YBCO thickness. Each link was thinned with 500 eV Ar ions impinging at 45° while the sample was cooled to ~230 K to avoid ion damage. The YBCO etch rate was calibrated to be ~12 nm/min by measuring the thickness of each milled sample with a Tencor profilometer. As noted later, there is an inherent uncertainty in the true YBCO thickness arising from a rough top surface of the YBCO and the tendency of the profilometer to measure the maximum height of the YBCO. After each milling step, the *V-I* characteristics were measured with a four-point configuration at 77 K for magnetic fields up to the irreversibility field applied perpendicular to the film surface. To prevent excessive heating at currents higher than 100 mA, a square current pulse of 50 ms duration with a 30 ms voltage read window was used. $J_c$ values were determined at a 1 μV/cm electric field criterion. Resistivity as a function of temperature, $\rho(T)$, was measured during sample cool down and $T_c$ was defined as the onset of zero resistance (~1% of normal state resistance). X-ray diffraction measurements were made to determine the global texture of the YBCO layers. In-plane texture was determined by the FWHM of the in-plane, off-axis ($\phi$) scan of the (103) YBCO peak, while the out-of-plane texture was measured from the rocking curve of the (005) YBCO peak.

## III. RESULTS

The key properties of the three samples at full thickness are shown in Table I. The highest $I_c^* = 313$ A/cm-width at $J_c$(0T,77K) = 2.5 MA/cm$^2$ was obtained in the 1.3 μm thick film (CC315), while the better textured 1.0 μm thick film (CC270) had $I_c^* = 272$ A/cm and the highest $J_c$ value of 2.8 MA/cm$^2$. $I_c^*$ of the less textured thinner film was somewhat lower; 128 A/cm and $J_c = 1.2$ MA/cm$^2$ due to its more misoriented grain boundary network and slightly different YBCO growth conditions [10]. The less textured sample (CC130) had in-plane full-width-half maximum (FWHM) ($\Delta\phi$) and out-of-plane FWHM ($\Delta\omega$) of 6.5° and 6.6°, as compared to values of ~5.4° and ~3.7° for the two higher $J_c$ CCs.

Figure 1a shows critical current per unit width $I_c^*$(77K, self-field) as a function of thickness $t$. The three data sets show smoothly increasing $I_c^*(t)$ curves with increasing $t$. For the two better textured samples (CC270/CC315), $I_c^*$ increases linearly up to ~ 0.2 μm at a steep slope of 390 Acm$^{-1}$μm$^{-1}$, while $I_c^*$ increases less steeply at 190 Acm$^{-1}$μm$^{-1}$ for the worse textured sample (CC130). The slopes decline with increasing $t$ to less than ~160-180 Acm$^{-1}$μm$^{-1}$ for CC270/CC315 and ~120 Acm$^{-1}$μm$^{-1}$ for CC130 near the film surface. However, all three curves yield zero current at non-zero thicknesses of 80, 50, and 150 nm, respectively. Figure 1b plots $I_c^*$ as a function of the thickness renormalized by subtracting the non-zero intercepts at which $I_c^*$ in Fig. 1a vanishes. Such renormalization of $t$ is applied throughout this paper. It corrects for uncertain effects due to the influence of the rough top surface and porosity, as discussed later. CC315 and CC270, which have essentially identical texture, show nice overlap in the renormalized plot of Fig. 1b.

Figure 2a shows the average (i.e. using $I_c$ averaged over the full residual thickness of the YBCO) $J_{c,avg}$(77K, self-field) as a function of $t$. It is striking that the two better textured samples reach a high $J_c$ value of ~4.0 MA/cm$^2$ near the bottom of the film and show a gradual and rather linear decrease to ~2.7 MA/cm$^2$ with increasing $t$. The less-textured sample, CC130, shows $J_c$ decreasing from ~1.8 MA/cm$^2$ to ~1.2 MA/cm$^2$ with increasing $t$. The incremental $J_c$ data are shown in Figure 2b as a function of distance from substrate. The two better textured samples exhibit a $J_c$ value of ~4.0 MA/cm$^2$ near the bottom and show a rather greater decline to ~ 1.6 MA/cm$^2$ at full thickness than do the average values. The CC130 shows more scattered data, but the tendency is the same with the better samples. When the incremental $J_c$ is plotted as in Fig. 2b, there is somewhat more scatter but the trends remain the same.

Figures 3a-f show a more detailed data set for the average and incremental in-field $J_c(t)$ for CC270 and CC315. The functional dependence is essentially identical to that seen at self field, showing that the thickness dependence is practically unaffected by the magnetic field, whether below or above the field (~2 T at 77 K) at which grain boundaries cease to limit the magnitude of $J_c$ of these films [10].

Figure 4a shows the room temperature resistivity, $\rho_{300K}$ as a function of $t$. The 3 samples have rather high $\rho_{300K}$ of 365 – 380 μΩcm at their full thickness, compared to the typical 200 – 250 μΩcm of high-quality, dense YBCO films [32-34]. The room temperature resistivity shows a monotonic decrease to ~300 μΩcm, as the YBCO layer is thinned to ~0.15 μm, the values being ~ 220-270 μΩcm at small thicknesses < 0.15 μm. The asymptotic $\rho$ value at zero thickness is ~ 200 μΩcm, a value expected for well oxygenated, dense YBCO films. We only see such a dense structure very near the interface, as described later. Insets to Fig. 4a show the local volume fraction of pores for CC270 and CC315 calculated using an effective medium theory (EMT), as detailed in the discussion, using a room temperature resistivity $\rho_{300K}$ appropriate for dense high quality YBCO films of 200 μΩcm. Figure 4b indicates the room temperature resistivity, $\rho_{300K}$ as a function of $t$, when calculated from the total measured thickness of Fig. 1a. Without the renormalization incited by the non-zero intercepts, the $\rho_{300K}$ values increase rapidly below ~0.2-0.3 μm to physically unreasonable values, which are quite inconsistent with the rather uniform depth profiled superconducting properties shown in Figures 5 and 6.

Figures 5a-c show the $\rho(T)$ curves normalized to $\rho_{300K}$. An important feature is that the curve shapes are independent of thickness for each sample, demonstrating both that the oxygenation state is identical through thickness and that repeated ion milling did not deoxygenate the film. The curves of CC130 and CC270 show similar slight upward curvatures and negative intercepts with the $\rho$ axis for the linear extrapolation of $\rho(T)$, which indicates that these two films are slightly oxygen over-doped [35], consistent with their lower $T_c$ values. The curve shape of CC315 shows almost linear $\rho(T)$ and the highest $T_c$, consistent with this sample being close to optimum doping. Figures 5d-f indicate that the $T_c$ transition behavior is also independent of thickness, both with respect to $T_c$ and the $T_c$ transition width $\Delta T_c$. Only a relatively small $T_c$ variation of 1-2 K is observed for each film, as shown in Figure 5g. This result also supports our conclusion that the oxygenation state of each film is independent of thickness.

Figure 6a-c shows the $J_c(H)$ curves, irreversibility field $H^*$ and the normalized bulk pinning force $(F_p/F_{p,max})$ for CC315 as a function of reduced field $H/H^*$, where the irreversibility field $H^*$ is defined at $J_c$ =100 A/cm$^2$ or by extrapolation of $F_p(H)$ to zero. Although the magnitude of $J_c(H)$ increases somewhat at smaller $t$, it is striking that $H^*$(77K) for all 3 samples and the shapes of the pinning force curves are also quite independent of thickness, indicating that the dominant vortex pinning mechanisms are also independent of film thickness.

Figure 7 shows plan-view SEM images of the original top surface of the 1 μm CC130 and of the milled surfaces at thicknesses of 0.78, 0.38, and 0.15 μm. Evidently, the original surface is quite rough and porous, but this roughness and porosity decreases continuously towards the substrate. The porosity variation is qualitatively consistent with the $\rho_{300K}$ vs. $t$ results and local porosity of inset of Fig. 4 if one assumes that the resistivity at any $t$ is controlled by the local porosity-controlled cross-sectional area.

Figure 8a shows a longitudinal section TEM image of CC270 that gives a broad overview of the through thickness porosity, while Figures 8b and 8c show higher resolution images of the same CC. Porosity with a scale in the range 0.2-0.5 μm which steadily increases towards the top of the YBCO is evident, as is the variability of porosity from place to place. The rather uneven top YBCO surface where it contacts the silver cap layer has a height variation of up to 0.2 μm. The film also shows many planar stacking faults parallel to the YBCO *ab* planes. These faults are typically 30 – 50 nm apart along the *c*-axis, are of high density, and distributed rather evenly through the film thickness. The laminar and porous microstructure is in strong contrast to the columnar and dense microstructure of PLD conductors.

Figure 9 shows two plan-view TEM images of the CC270 sample which also indicate porosity change through thickness. Fig. 9a, taken ~ 0.2 μm below the top surface, shows high porosity. In contrast, the image taken near the bottom [Fig. 9b] shows much less porosity. The pores present in Fig 9a vary in scale from 80 – 400 nm.

## IV. DISCUSSION

A striking feature of these MOD CCs is their high average $J_c$ ~ 3 MA/cm$^2$ and high $I_c$ values (~ 300 A/cm) at thicknesses of ~ 1 μm, even though these $J_c$ or $I_c$ values are definitely limited by higher angle grain boundaries [10] and by their significant porosity. They also have a $J_c(t)$ dependence much weaker than most PLD [2-8] and earlier standard baseline PVD-BaF$_2$ processed conductors [22,27]. Figs. 5 and 6 show that the present MOD films have substantially uniform local properties, as indicated by the thickness independence of $T_c$, $\Delta T_c$, the shape of the bulk flux pinning curve $F_p(H)$, the irreversibility field $H_{irr}$ and the resistivity curve $\rho(T)$ shape. Consistent with these observations, microscopy shows no significant change of microstructure with thickness, except for a marked increase in porosity with increasing thickness. Moreover, these samples also show marked changes in the *magnitudes* of room temperature resistivity $\rho_{300K}$ and both the average $J_c$ and the incremental $J_c$ that plausibly correlate to a thickness-dependent *area degradation*. From our effective medium calculations of the resistivity discussed below, we conclude that the local vortex-pinning-determined $J_c$ is nearly independent of thickness but that the average critical current density $J_{c,avg}(t)= I_c/A$ defined by the whole YBCO cross-section $A$ depends on $t$ precisely because the thickness dependence of porosity causes the current-carrying cross-section to vary. The striking paradox of such a conclusion is that it is these porous and somewhat rough MOD films that show higher $J_c$ and weaker $J_c(t)$ than the denser PLD films because vortex pinning is enhanced by small nanoscale pores, even as larger or locally clustered pores may obstruct current flow. As coated conductor process scale up continues and the ultimate limits of coated conductor technology are explored, the mechanisms which control the thickness dependence of the critical current density demand a better understanding of this apparent paradox. We focus our discussion on a comparison of the magnitude of $J_c$ and of $J_c(t)$ of our MOD films to the behavior exhibited by coated conductor grown by other methods.

The irregular surface and porosity of MOD films shown in Figs. 7-9 is a direct consequence of the coating technique, which lays down a precise quantity of YBCO precursor over a defined substrate area. The decomposition of the carboxylate precursors and subsequent evolution of gas-phase HF, produce ~50% reduction in original YBCO precursor thickness [28]. Local variations of porosity produced by these processes result in an unavoidable variation in the local cross-section of YBCO. The profilometer may also over-estimate the thickness $t$ because its 12.5 µm tip radius is much larger than the local thickness variation due to rough surface ~ 0.1 µm (Fig. 8). After ion milling, top-surface roughness does propagate to smaller thicknesses with some smoothing and with a cross-sectional uncertainty compounded by porosity inside the film. We believe that these factors contribute to the non-zero intercept seen in Fig. 1a. Fig. 4b also indicates the physically unreasonable $\rho_{300K}$ values when the total measured (over-estimated) thickness is taken into account. Indeed, the $\rho_{300K}(t)$ is inconsistent with other results, including uniform superconducting properties through thickness and observed porosity change. Actually, Fig. 8a suggests a local thickness of ~ 1.1 µm for CC270 with peak-to-valley variation > ~0.1 µm. It appears that the thickness of this type of MOD conductor cannot be determined to better than ~ ± 10 %. Collectively these measurements lead to an over-estimate of film thickness, thus leading to a small under-estimate of the $J_c$ value derived from the whole cross-section.

The effect of porosity on $J_c$ can manifest itself in different ways. Large scale porosity and surface roughness on the scale larger than the London penetration depth $\lambda(77K)$ ~0.4 µm mainly obstruct current flow. However, close examination of Figs. 7-9 shows that pores have multiple scales, many smaller than 0.4 µm. Such nanoscale porosity can enhance flux pinning by either caging the vortex screening currents (magnetic pinning) or by vortex-core interactions with small pores (vortex core pinning). An additional positive feature of the porosity is that it should enhance full oxygenation of the film, a property which is consistent with the through-thickness uniformity of $T_c$, $\Delta T_c$ $F_p(H)$, $H_{irr}$ and the normalized shape of $\rho(T)$ [Figs. 4-6].

As noted earlier, extensive $J_c(t)$ data sets for PLD films [2-8], generally show $t^{-1/2}$-like behavior at small t, both for films with clear microstructural degradations at top or bottom [2] or recently grown films without such degradations [7]. The recent study of Emergo *et al.* [8] with miscut substrates which produce characteristic porosity distributions also shows a similar $t^{-1/2}$ dependence, suggestive of a 2D to 3D pinning crossover behavior at $t_c$ ~0.5-1 μm. These latter PLD samples have no clearly identified pinning structures and have $J_c$ *(77K,* self field) ~1 MA/cm$^2$ at t = 1-2 μm. To see if any kind of surface pinning, can account for the observed $J_c$ values, we first estimate $J_c$ in the strong surface pinning limit, for which the ends of vortices are fixed only by surface irregularities or other defects, for example, misfit dislocations as suggested by Foltyn et al [17]. In this case $J_c$ is determined by depinning of a critical vortex loop whose diameter is equal to the separation d between the fixed ends of the vortex segment [36], as shown in Fig. 11a :

$$J_c = \frac{\phi_0}{2\pi\mu_0 \lambda_a \lambda_c d} \ln \frac{d}{\xi_c} \qquad (1)$$

Here $\phi_0$ is the flux quantum, $\lambda_a$ and $\lambda_c$ are the London penetration depths in the *ab* plane and along the *c*-axis, respectively, and $\xi_c$ is the coherence length along the *c*-axis. For pure surface or interface pinning, d equals the film thickness, in which case Eq. (1) yields $J_c \approx 0.22$ MA/cm$^2$ for a film of thickness 1 μm at 77K if we take $\lambda_a = 0.4$ μm, $\lambda_c = 2$ μm, and $\xi_c = 1$ nm at 77 K. Thus, surface pinning, no matter how strong, is insufficient, to account for the typical $J_c$ ~ 1-4 MA/cm$^2$ found in YBCO coated conductors and films. Such high $J_c$ values can only result from dense, strong bulk pinning. If we now use Eq.(1) to estimate what average spacing d between such strong pins is needed to provide $J_c(77K) = 4$ MA/cm$^2$, we find that d ~ 30 nm is much smaller than the film thickness. This result is qualitatively consistent with earlier estimates of d by Hylton and Beasley [37]. Thus, vortex pinning defects in YBCO thin films have to be dense and strong to explain the current densities of 2-4 MA/cm$^2$ observed in all good quality films and coated conductors at 77K.

Another large, variable-thickness data set is for films grown by the *ex situ* PVD-BaF$_2$ process by Feenstra *et al.* [27], using slow-conversion ("baseline") and faster-conversion processes. The $I_c(t)$ data of the baseline process films support $J_c \propto t^{-1/2}$ very well ($I_c \sim t^{1/2}$), while the fast-process films show a linear $I_c(t)$ behavior, similar to our MOD films. For the baseline process, $J_c$ is as low as ~ 0.5 MA/cm$^2$ when the thickness is larger than 1 μm (~ $t_c$), which is not very different from the ~1 MA/cm$^2$ of the PLD films discussed above, if grain boundary effects that reduce the magnitude of $J_c$ in RABiTS-grown samples are accounted for. However, $J_c$ is much higher ~ 2.5 MA/cm$^2$ for the fast conversion-process RABiTS films that show the linear $I_c(t)$ behavior. The very different $J_c(t)$ behavior exhibited by different processing of the same precursors is another sign that microstructural effects are playing a decisive role in controlling the $J_c(t)$ behavior.

Taken together, these two data sets on PLD and on $BaF_2$ films show that real differences in both the magnitude of $J_c$ and in the $J_c(t)$ relationship exist for both *in situ* PLD films and in *ex situ* $BaF_2$ films. Thus, we wish to explore the way by which an optimized pinning microstructure can enhance the vortex-pinning $J_c$ magnitude, taking into account the fact that materials defects can also obstruct the current-carrying cross-section, and degrade the average $J_c$ deduced from dividing $I_c$ by the whole YBCO cross-section. In trying to understand the mechanisms underlying individual $J_c(t)$ data sets, several points should be addressed:

(1) Is the pinning microstructure independent of thickness?

(2) Is the current-carrying cross-section independent of thickness?

(3) Are grain boundary limitation effects present?

Turning to the first question, it seems that the pinning microstructure of our MOD films is independent of thickness, as judged from the thickness independence of $T_c$, $\Delta T_c$, the shape of the bulk flux pinning curve $F_p(H)$, the irreversibility field $H_{irr}$, and the normalized resistivity curve $\rho(T)$ shape. Microscopy also suggests no significant change of microstructure with thickness, except for porosity. In this respect these MOD samples appear to be very different from the PLD samples of Foltyn *et al.* [17] which exhibit very high interface $J_c$ up to ~ 7 $MA/cm^2$ that the authors associate with an interface dislocation array. The question of whether the current-carrying cross-section is independent of thickness is particularly relevant to our MOD conductors, for which increase of porosity is observed with increasing thickness, as shown in the cross-sectional TEM image of Fig. 8, the plan-view SEM images of Fig. 7 and the plan-view TEM images of Fig 9. Only in Fig 9b taken at the bottom of the YBCO layer is the porosity apparently absent. Such large-scale porosity must limit the connected cross-section, a result that is seldom seen for well-made PLD films. As for the third question of whether grain boundary cross-section limitation effects are present, an earlier study answers this question [10]. These conductors were made on highly textured RABiTS$^{TM}$, but even for the two better textured samples, major grain boundary network limitation effects are present below ~2T, as seen in our previous work [10]. CC270 and an MOD single crystal film had 77 K, self-field average $J_{c,avg}$ values of ~2.6 and ~5.3 $MA/cm^2$ respectively. This latter value of 5.3 $MA/cm^2$ unambiguously places the MOD-process films as strong vortex-pinning YBCO films, in spite of their significant porosity and their increased $\rho_{300K}$ values.

To estimate the effect of pores in reducing the current-carrying cross section $A_{eff}$, we use a conventional effective medium theory (EMT) [38], assuming that the sample contains pores of volume fraction 1 - $x$ embedded in a metallic matrix with anisotropic resistivity $\rho_a$ in the *ab* plane and $\rho_c$ along the *c*-axis. As shown in the Appendix, the anisotropic EMT gives the following expression for the measured in-plane resistivity $\rho$, and $A_{eff}$

$$\rho = \rho_a \frac{1-x_c}{x-x_c}, \qquad A_{eff} = A\frac{x-x_c}{1-x_c}, \qquad (2)$$

where $A$ is the geometrical cross-sectional area, and $x_c$ depends on the anisotropy ratio $\varepsilon = \rho_a/\rho_c < 1$:

$$x_c = \frac{1}{2(1-\varepsilon)}\left[1 - \frac{\varepsilon}{2\sqrt{1-\varepsilon}}\ln\frac{1+\sqrt{1+\varepsilon}}{1-\sqrt{1-\varepsilon}}\right] \qquad (3)$$

Eq. (2) shows that $\rho$ increases and $A_{eff}$ decreases as the volume fraction $x$ of the conductor decreases. The resistivity becomes infinite and $A_{eff}$ vanishes if $x$ is smaller than the EMT percolation threshold $x_c$, which varies between 1/2 for the highly anisotropic 2D case $\varepsilon \to 0$ and 1/3 for the isotropic 3D case, $\varepsilon \to 1$. For YBCO, $\varepsilon = \rho_a/\rho_c = 0.2$ and Eq. (3) yields $x_c = 0.42$. Next we calculate the current-carrying volume fraction $x$ of YBCO in our films, re-writing Eq. (2) in the form:

$$x = x_c + (1-x_c)\frac{\rho_a}{\rho} \qquad (4)$$

Taking $\rho_a \sim 200$ μΩcm (the asymptotic value at zero thickness in Fig.3) and $\rho = 380$ μΩcm, we obtain a connected cross-section x = 0.73. This result is in general accord with the microstructural images in Figs. 7-9. This total 27% pore volume obviously can considerably obstruct current flow, as well as providing some vortex pinning by the smaller pores. The substantial thickness variation of the normal state resistivity is clear evidence that significant obstruction of the normal-state current occurs, when supported by the lack of through-thickness change to the local superconducting properties. Using Eq. (2) to extract the EMT-current-carrying cross-section from the resistivity data in Fig. 4, we then calculated the thickness-dependent porosity which is shown in insets to Fig. 4 and then used these data to calculate a porosity-renormalized critical current density $J_c = I_c/A_{eff}$ for each $t$, as shown with open symbols in Figures 10a and 10b. Consistent with our assessment that the *local* properties do not change significantly through thickness, this porosity-renormalized $J_c$ (but not yet grain boundary renormalized) exhibits even less thickness dependence and even higher average $J_{c,avg}$ values 4–5 MA/cm$^2$. Although the incremental $J_c(t)$ data in Fig. 10b are more scattered than the full-thickness averages in Fig. 10a, the incremental $J_c(t)$ data now suggest that there is NO thickness dependence in the bottom half of the films. Beyond about 0.6 μm thickness, $J_c(t)$ then falls off from the plateau of ~4.5 MA/cm$^2$ to ~2 MA/cm$^2$ at ~1.1 μm. The basis for this change is not certain, although one possibility is that the porosity is smaller and closed towards the bottom of the film and thus more useful for flux pinning than the more open, crevasse-like, current-blocking porosity that appears at the top of the film.

The high $J_c$ values reported here may originate from multiple defect structures, including pores, twins, point defects, stacking faults, $Y_2O_3$ particles, as well as atomic-scale disorder. Civale et al. [30] have pointed out that there is a strong *ab* plane pinning peak in these MOD films, which they correlate to the significant volume fraction of ab planar defects observed clearly in Figs. 7b and 7c. Thus, we conclude that these MOD films are always in the strong 3D vortex pinning regime, for which $J_c$ should indeed be independent of thickness as suggested by Fig. 10b, because strong pins spaced by 10 – 100 nm subdivide vortices into short, nearly decoupled segments much smaller than the film thickness (t >> d). This conclusion is consistent both with the observed independence of $J_c$ on t and the estimate based on Eq. (1), which requires a mean pin spacing of order 30 nm. By contrast, the $t^{-1/2}$ dependence of $J_c$ seen in PLD films may indicate that vortex lines are pinned by many weaker defects which do not disrupt the continuity of the vortex lines.

These two different types of pinning behavior are illustrated by Fig. 11. Fig. 11a shows that vortex lines can be chopped into discontinuous segments by strong pinning defects like pores or $Y_2O_3$ particles, aided perhaps by the planar defects seen in Fig. 8, for which $J_c$ can be estimated from Eq. (1). This case produces the thickness-independent $J_c$ of the MOD films studied in this work. By contrast, Fig. 11b shows the case of a weaker, although still dense array of pins, which perturb the vortex lines but do not make them discontinuous. Here $J_c$ can be estimated using the 2D collective pinning theory for a rigid vortex line in a thin film where the bending distortions of the vortex are negligible [14,15]. If such pins are distributed randomly with mean density $n_i = 1/d^3$ and each pin exerts a pinning force $f_p$ with interaction radius $r_p$, the net pinning force on the vortex line equals $F_p = f_p N^{1/2}$ where $N = \pi t r_p^2 / d^3$ is the total number of pins in the cylinder of radius $r_p$ and height t. Here the factor $N^{1/2}$ reflects the fact that a vortex adjusts its position to find a local minimum in the random potential produced by statistical fluctuations of all the defects. Balancing $F_p$ against the total Lorentz force, $F_L = \phi_0 t J_c$, we obtain

$$J_c \cong \frac{\pi^{1/2} f_p r_p}{\phi_0 d^{3/2} t^{1/2}} \approx \frac{J_d r_p \xi}{d^{3/2} t^{1/2}}, \qquad (5)$$

where the second relation is estimated for strong pins for which we took the maximum pinning force $f_p \sim \phi_0 \xi J_d$, where $J_d$ is the in-plane depairing current density. Eq. (5) gives the inverse square root dependence $J_c \propto t^{-1/2}$ observed for typical PLD films. Eq. (5) can also be used to estimate the mean pin spacing d, which would provide typical self-field values of $J_c \sim 2$ MA/cm$^2$ in a 0.5 μm thick PLD film at 77K. Taking $\xi = 4$nm, t = 0.5 μm, $J_d = 40$ MA/cm$^2$ and $r_p = 20$ nm, we obtain the mean spacing d $\sim \xi (J_d r_p / J_c \xi)^{2/3} (\xi/t)^{1/3}$ $\sim 4.3\xi \sim 17$ nm of the order of the interaction radius, which may represent a characteristic size of pinning nanoprecipitates. This estimate suggests that the mean pin spacing in PLD films is shorter than in MOD films. However MOD films can have larger, sparser but stronger pins and a consequently much weaker thickness dependence of $J_c(t)$ as compared to PLD films. Of course thee are many types of PLD films that now incorporate pinning centers such as $YBa_2CuO_5$ [39] and $BaZrO_3$ [40] that are also capable of putting PLD films into this same 3D pinning regime, but explicit demonstration of this point has not yet been made. The remarkable result of our study of these MOD films is that they naturally put themselves into this strong pinning 3D limit, even without added rare-earth nanodots [41]. Given the twice larger average $J_c$ values of the single crystal version of the RABiTS conductors [10], it seems likely that the true local vortex pinning $J_c$ is of order twice the values in Fig. 10, making them ~ 8 MA/cm$^2$, but more advanced studies are needed to define the source and magnitude of this strong pinning.

## V. SUMMARY

We have made detailed studies of the thickness dependence of the superconducting properties and the normal state resistivity $\rho$ after successive ion milling of ~ 1 μm thick YBa$_2$Cu$_3$O$_{7-x}$ films made by an *ex situ* metal-organic deposition process on Ni-W rolling-assisted biaxially textured substrates. Contrary to much recent data, mostly on *ex situ* films, showing strong depression of the critical current density $J_c$ with increasing film thickness $t$, our films exhibit only a weak dependence of $J_c$ on increasing $t$. The two better textured samples had full cross-section average $J_c$(77K,0T) ~ 4 MA/cm$^2$ near the buffer layer interface and ~ 3 MA/cm$^2$ at full thickness, which shows that strong vortex pinning is operating, despite significant current blocking limitations due to porosity and grain boundaries. We conclude that these MOD films exhibit strong three-dimensional pinning which produces a local $J_c$ which is independent of thickness. Thickness-dependent changes in microstructure, mostly porosity which increases with thickness, are found to be responsible for the observed weak thickness dependence of the average $J_{c,avg} = I_c/A.$.

## ACKNOWLEDGEMENTS

This work was supported by the Air Force Office of Scientific Research (AFOSR) MURI Award and by the DOE Office of Electric Delivery and Energy Reliability.

## APPENDIX

To calculate the effective resistivity $\rho$ of a conductor with spherical pores we calculate the electric field distribution $\mathbf{E}(\mathbf{r}) = -\nabla\varphi$ in a medium with a randomly inhomogeneous conductivity $\sigma(\mathbf{r})$. The equation for the electric potential $\varphi$ is:

$$\frac{\partial}{\partial x}\sigma_a\frac{\partial\varphi}{\partial x} + \frac{\partial}{\partial y}\sigma_a\frac{\partial\varphi}{\partial y} + \frac{\partial}{\partial z}\sigma_c\frac{\partial\varphi}{\partial z} = 0 \qquad (A1)$$

where $\sigma_a(r) = 1/\rho_a$ and $\sigma_c(r) = 1/\rho_c$ are the normal state conductivities along the ab plane and the c-axis, respectively between the pores and $\sigma = 0$ in the pores. Next, we transform the z coordinate along the c-axis, $z \rightarrow (\sigma_c/\sigma_a)^{1/2}z$ so that $\sigma$ in Eq. (A1) cancels. In the new coordinates each pore becomes an ellipsoid stretched along the c-axis with the in-plane to the out-of-plane semi-axis ratio $\varepsilon = (\sigma_c/\sigma_a)^{1/2}$. In the EMT models the electric field distribution in a heterogeneous conductor results from local electric fields produced by randomly-distributed ellipsoids with a different local in-plane resistivities $\rho_i$, embedded in a uniform effective medium with the exact global resistivity ρ. The electric field in the ellipsoid is uniform and independent of its size [42]:

$$E_i = \frac{E\rho_i}{(1-n_x)\rho_i + n_x\rho} \quad (A2)$$

where E is the mean electric field, and $n_x$ is the depolarization factor of the ellipsoid along the x-axis:

$$n_x = \frac{1}{2(1-\varepsilon)}\left[1 - \frac{\varepsilon}{2\sqrt{1-\varepsilon}}\ln\frac{1+\sqrt{1+\varepsilon}}{1-\sqrt{1-\varepsilon}}\right] \quad (A3)$$

Averaging Eq. A(2) with the distribution function $F(\rho_i)$ gives the main EMT self-consistency equation for the global resistivity ρ:

$$1 = \int_0^\infty \frac{F(\rho_i)\rho_i d\rho_i}{(1-n_x)\rho_i + n_x\rho} \quad (A4)$$

If a heterogeneous conductor has a volume fraction 1 – x of pores with $\rho_i = \infty$ in a uniform metallic matrix with $\rho_i = \rho_a$, Eq. (A4) becomes

$$1 = \frac{x\rho_a}{(1-n_x)\rho_a + n_x\rho} + \frac{1-x}{1-n_x} \quad (A5)$$

Solving Eq. (A5) for ρ, we obtain Eq. (1) with the percolation threshold, $x_c = n_x$.

## References

1. D. Larbalestier, A. Gurevich, D. M. Feldmann, et al., Nature **414**, 368 (2001).
2. S. R. Foltyn, P. Tiwari, R. C. Dye, et al., Appl Phys Lett **63**, 1848 (1993).
3. S. R. Foltyn, Q. X. Jia, P. N. Arendt, et al., Appl Phys Lett **75**, 3692 (1999).
4. X. D. Wu, S. R. Foltyn, P. Arendt, et al., IEEE Trans Appl Supercon **5**, 2001 (1995).
5. B. W. Kang, A. Goyal, D. R. Lee, et al., J Mater Res **17**, 1750 (2002).
6. J. R. Groves, P. N. Arendt, S. R. Foltyn, et al., J Mater Res **16**, 2175 (2001).
7. S. R. Foltyn, P. N. Arendt, Q. X. Jia, et al., Appl Phys Lett **82**, 4519 (2003).
8. R. L. S. Emergo, J. Z. Wu, T. Aytug, et al., Appl Phys Lett **85**, 618 (2004).
9. K. Develos-Bagarinao, H. Yamasaki, J. C. Nie, et al., Supercond Sci Tech **18**, 667 (2005).
10. S. I. Kim, D. M. Feldmann, D. T. Verebelyi, et al., Phys Rev B **71**, 104501 (2005).
11. T. G. Holesinger, P. N. Arendt, R. Feenstra, et al., J Mater Res **20**, 1216 (2005).
12. D. M. Feldmann, D. C. Larbalestier, T. Holesinger, et al., J Mater Res **20**, 2012 (2005).
13. C. J. van der Beek, M. Konczykowski, A. Abal'oshev, et al., Phys Rev B **66**, 024523 (2002).
14. P. H. Kes and C. C. Tsuei, Phys Rev B **28**, 5126 (1983).

15  A. Gurevich, Presentation at http://www.energetics.com/meetings/supercon04/pdfs/presentations/f_uw_coated_conductor_peer_rev_04final.pdf.
16  L. Civale, B. Maiorov, A. Serquis, et al., Appl Phys Lett **84**, 2121 (2004).
17  S. R. Foltyn, H. Wang, L. Civale, et al., Appl Phys Lett **87** (2005).
18  L. Fernandez, B. Holzapfel, F. Schindler, et al., Phys Rev B **67**, 052503 (2003).
19  T. Ohnishi, R. H. Hammond, and W. Jo, J Mater Res **19**, 977 (2004).
20  W. Wong-Ng, I. Levin, R. Feenstra, et al., Supercond Sci Tech **17**, S548 (2004).
21  W. Wong-Ng, L. P. Cook, J. Suh, et al., Supercond Sci Tech **18**, 442 (2005).
22  D. M. Feldmann, D. C. Larbalestier, R. Feenstra, et al., Appl Phys Lett **83**, 3951 (2003).
23  S. I. Kim, et. al, Manuscript in prep.
24  V. F. Solovyov, H. J. Wiesmann, L. J. Wu, et al., Ieee T Appl Supercon **9**, 1467 (1999).
25  J. A. Smith, M. J. Cima, and N. Sonnenberg, Ieee T Appl Supercon **9**, 1531 (1999).
26  M. W. Rupich, D. T. Verebelyi, W. Zhang, et al., Mrs Bull **29**, 572 (2004).
27  R. Feenstra, A. A. Gapud, F. A. List, et al., Ieee T Appl Supercon **15**, 2803 (2005).
28  M. W. Rupich, W. Zhang, X. Li, et al., Physica C **412-14**, 877 (2004).
29  M. W. Rupich, U. Schoop, D. T. Verebelyi, et al., IEEE Trans Appl Supercon **13**, 2458 (2003).
30  L. Civale, B. Maiorov, A. Serquis, et al., Physica C **412-14**, 976 (2004).
31  T. G. Holesinger, D. M. Feldmann, and R. Feenstra, Presentation at http://www.energetics.com/meetings/supercon04/pdfs/presentations/h_lanl_ornl_uw.pdf.
32  C. B. Eom, A. F. Marshall, Y. Suzuki, et al., Phys Rev B **46**, 11902 (1992).
33  A. C. Westerheim, A. C. Anderson, D. E. Oates, et al., J Appl Phys **75**, 393 (1994).
34  O. Castano, A. Cavallaro, A. Palau, et al., Supercond Sci Tech **16**, 45 (2003).
35  R. Feenstra, D. K. Christen, C. E. Klabunde, et al., Phys Rev B **45**, 7555 (1992).
36  E. H. Brandt, Phys Rev Lett **69**, 1105 (1992).
37  T. L. Hylton and M. R. Beasley, Phys Rev B **41**, 11669 (1990).
38  J. P. Clerc, G. Giraud, J. M. Laugier, et al., Adv Phys **39**, 191 (1990).
39  T. Haugan, P. N. Barnes, R. Wheeler, et al., Nature **430**, 867 (2004).
40  J. L. Macmanus-Driscoll, S. R. Foltyn, Q. X. Jia, et al., Nat Mater **3**, 439 (2004).
41  U. Schoop, M. W. Rupich, C. Thieme, et al., IEEE T Appl Supercon **15**, 2611 (2005).
42  J. D. Jackson, "Classical Electrodynamics", John Wiley & Sons, New York (1998).

**Table I. Key properties of the three coated conductor samples**

|                | CC130   | CC270   | CC315   |
|----------------|---------|---------|---------|
| Full thickness | 1.1 μm  | 1.0 μm  | 1.3 μm  |
| $\Delta\phi$ (YBCO) | 6.5° | 5.5° | 5.4° |
| $\Delta\omega$ (YBCO) | 6.6° | 3.8° | 3.7° |
| $J_c$(77K, self-field) | 1.2 MA/cm$^2$ | 2.8 MA/cm$^2$ | 2.5 MA/cm$^2$ |
| $I_c^*$(77K, self-field) per cm width | 128 A/cm | 272 A/cm | 313 A/cm |
| $T_c$ | 88.7 K | 91.3 K | 92.9 K |

**Figure Caption**

FIG. 1. (a) The critical current per unit width at $I_c^*$ (77K,self-field) vs. the measured thickness for three different coated conductors. $I_c^*$ goes to zero at non-zero thicknesses of 80, 50, and 150 nm, respectively. $I_c^*$ initially increases linearly with a slope of 390 Acm$^{-1}$μm$^{-1}$ for CC315 and CC270 and 190 Acm$^{-1}$μm$^{-1}$ for CC130, with the slope becoming smaller at larger thickness. (b) $I_c^*$ (77K,self-field) vs. the renormalized thickness, which subtracts the non-zero intercepts. CC315 and CC270, which have similar texture properties, now show excellent agreement.

FIG. 2. (a) The average $J_c$(77K,self-field) vs. thickness after renormalizing the thickness to subtract off the non-zero intercept in Figure 1(a). $J_c$(t) shows rather slow reduction with increasing thickness. CC315 and CC270 exhibit very high $J_c$ of ~ 4 MA/cm$^2$ at small thickness, while showing ~ 3 MA/cm$^2$ at full thicknesses. The CC130 shows about half these $J_c$ values due to its worse texture. (b) Incremental $J_c$ data shows rather steep linear decrease to ~1.6-1.7 MA/cm$^2$ near top surface for the better textured samples.

FIG 3. In-field $J_c$(t) data for CC270 and CC315. The $J_c$(t) or incremental $J_c$(t) data in fields is more or less identical to that at self-field.

FIG. 4. (a) The room temperature resistivity, $\rho_{300K}$, as a function of thickness. The $\rho_{300K}$ values at full thickness are high, reaching ~ 365 - 380 μΩcm for all samples. However, the $\rho_{300K}$ values show a general linear decline with decreasing thickness down to ~0.15 μm, and steeper decreasing below to a limiting value of ~ 200 μΩcm at zero thickness. This value is typical for dense, well-oxygenated YBCO. The insets show the local porosity of CC270 and CC315 calculated by effective medium theory. (b) $\rho_{300K}$ as a function of $t$, when calculated from measured thickness (which is shown in Fig. 1a). The $\rho_{300K}$ values increase rapidly below ~0.2-0.3 μm to physically unreasonable values.

FIG. 5. The plots show the $\rho(T)$ curves normalized to $\rho_{300K}$ for (a)(d) CC130, (b)(e) CC270, and (c)(f) CC315. Remarkably, the curve shapes are identical for all thicknesses for each sample. The curves of CC130 and CC270 show slight upward curvature and a slight negative linear intercept typical of overdoped YBCO, while CC315 shows almost linear $\rho(T)$. All samples are well oxygenated through their whole thickness. (g) $T_c$ as a function of thickness, where $T_c$ is determined at the onset of $\rho=0$. The relatively small $T_c$ variation within 1-2 K is seen for each sample.

FIG. 6. (a) The critical current density as a function of applied field $J_c(H)$ for CC315. (b) Normalized bulk pinning force ($F_p/F_{p,max}$) vs. reduced field ($H/H_{extrapolate}$) for CC315. $H_{extrapolate}$ is defined by extrapolation of linear sections of the high-field curves, as indicated. Neither the shapes of the pinning force curves or $J_c(H)$ change as a function of thickness. (c) The irreversibility field –$H_{irr}$ at $J_c=100A/cm^2$ and $H_{extrapolate}$ in $F_p(H)$ plot - as a function of thickness is shown. The irreversibility field is constant as a function of thickness.

FIG. 7. Plan view SEM images of (a)(e) the original surface of CC130 at 1 μm thickness and after milling the CC130 sample to (b)(f) 0.78, (c)(g) 0.38, and (d)(h) 0.15 μm thickness. The rough and porous nature of the film surface is quite evident, as is the decline in porosity near the substrate. The porosity variation is qualitatively consistent with the resistivity results of Fig.3.

FIG. 8. Cross-sectional TEM images of the CC270 sample; (a) a broad overview of the porosity, (b)(c) in higher resolution. Porosity with a scale in the range 0.2-0.5 μm increases towards the top of the YBCO. The rather uneven top YBCO surface where it contacts the silver cap layer has a height variation of up to 0.2 μm. The film also shows many planar stacking faults parallel to the YBCO *ab* planes, which are typically 30 – 50 nm apart along the *c*-axis, are of high density, and distributed rather evenly through the film thickness. The laminar and porous microstructure is in strong contrast to the columnar and dense microstructure of PLD conductors.

FIG. 9. Plan-view TEM images of the CC270 sample. (a) a image taken ∼ 0.2 μm below the top surface and shows high porosity, indicating porosity in various scales. (b) a image taken near the bottom shows much less porosity.

FIG. 10. (a) Porosity-renormalized average $J_{c,avg}(t)$ using effective medium theory (EMT) is shown with open symbols to take account the porosity change through thickness. (b) Porosity-renormalized incremental $J_c(t)$.

FIG.11. (a) Illustration of the way that vortex lines can be chopped into short discontinuous segments by strong pinning defects which enforce depinning by the bending of a vortex segments in a critically sized vortex loop, whose diameter is equal to the separation d between pins . This case corresponds to the present MOD films. (b) Illustration of continuous vortex lines in a denser array of weaker pins, which leads to the inverse square root dependence on thickness seen in many normally made PLD films.

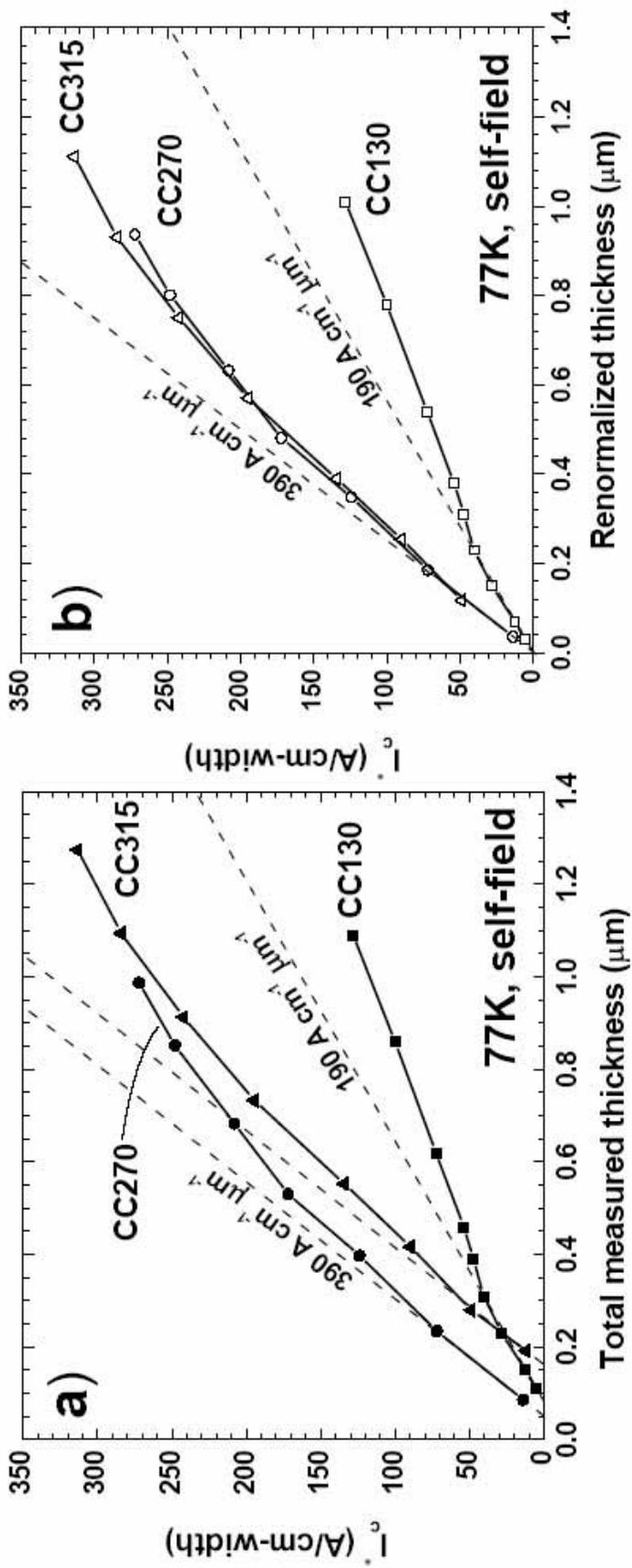

Fig 1. S.I. Kim et al.

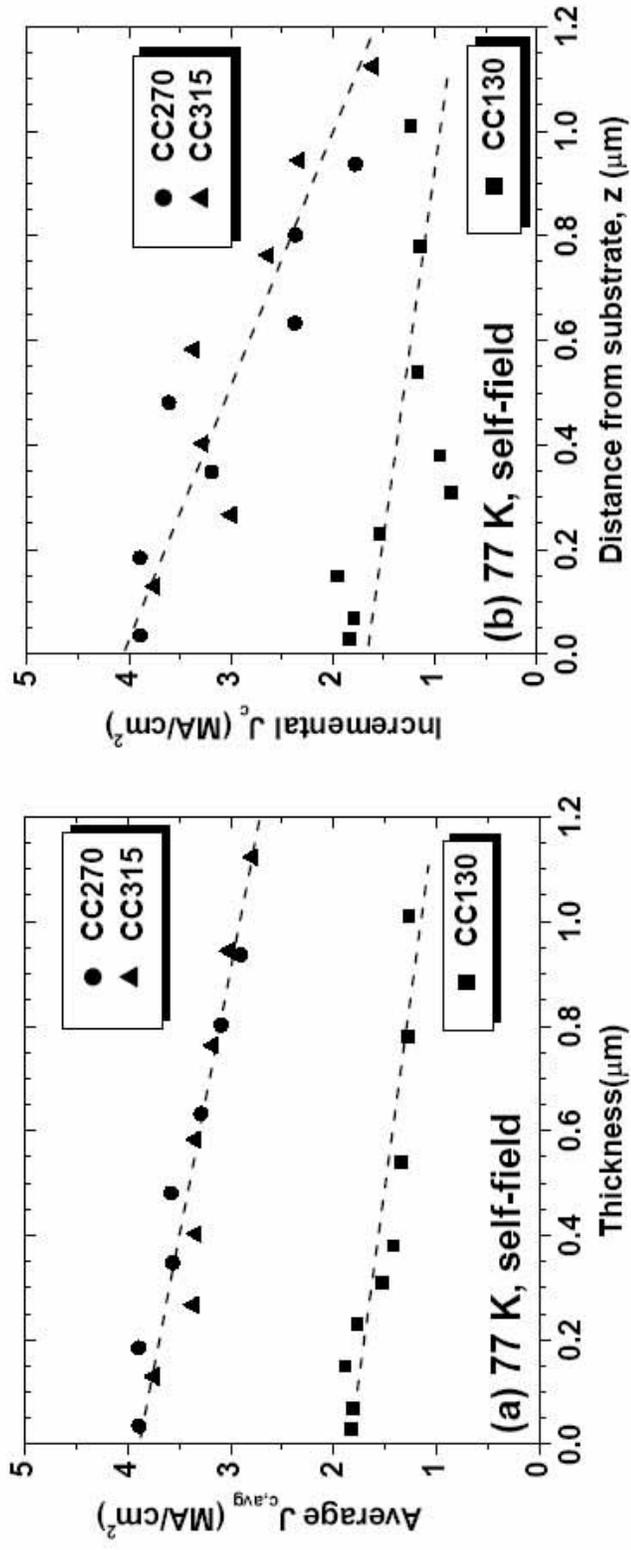

Fig 2. S.I. Kim et al.

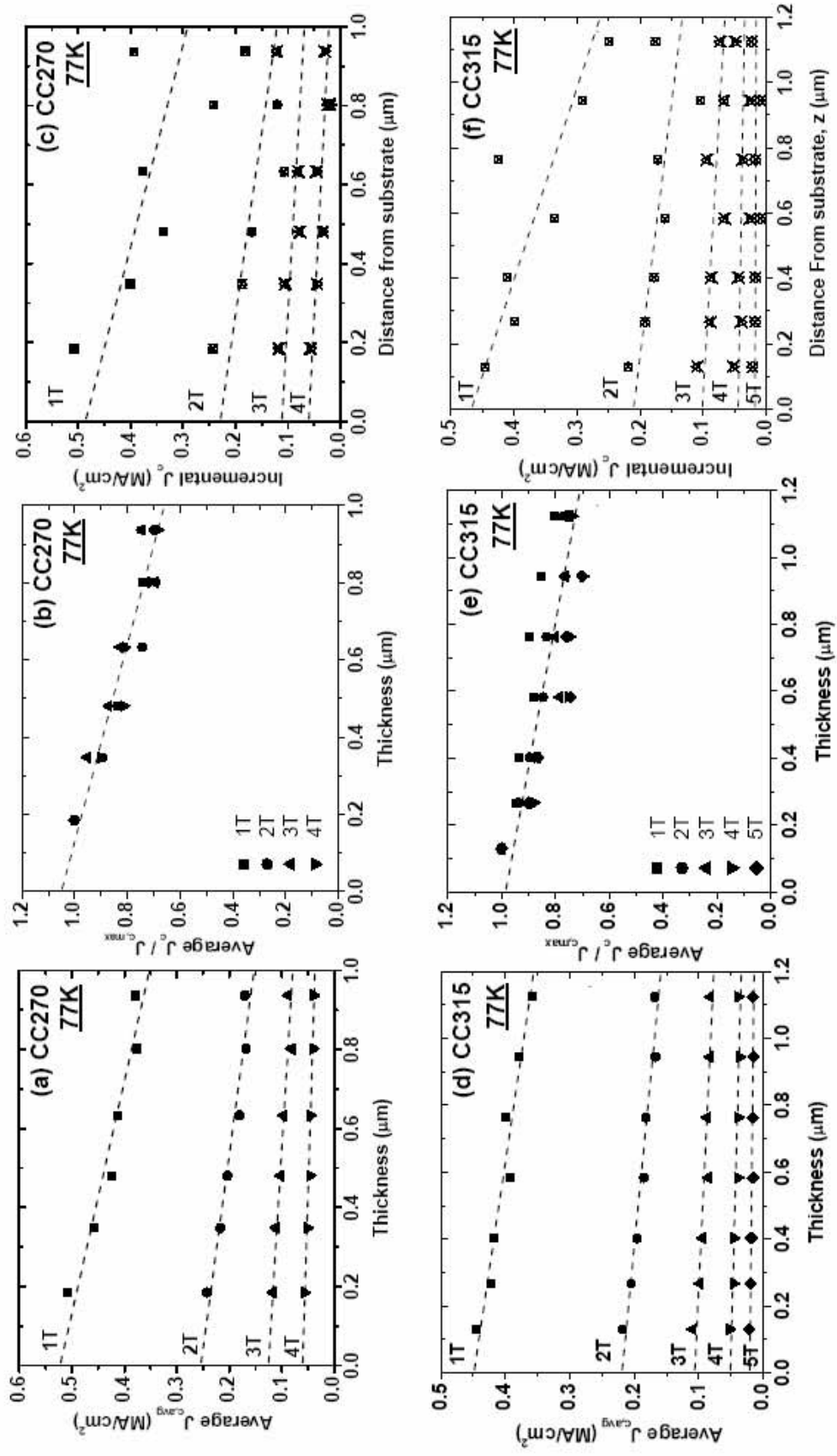

Fig 3. S.I. Kim et al.

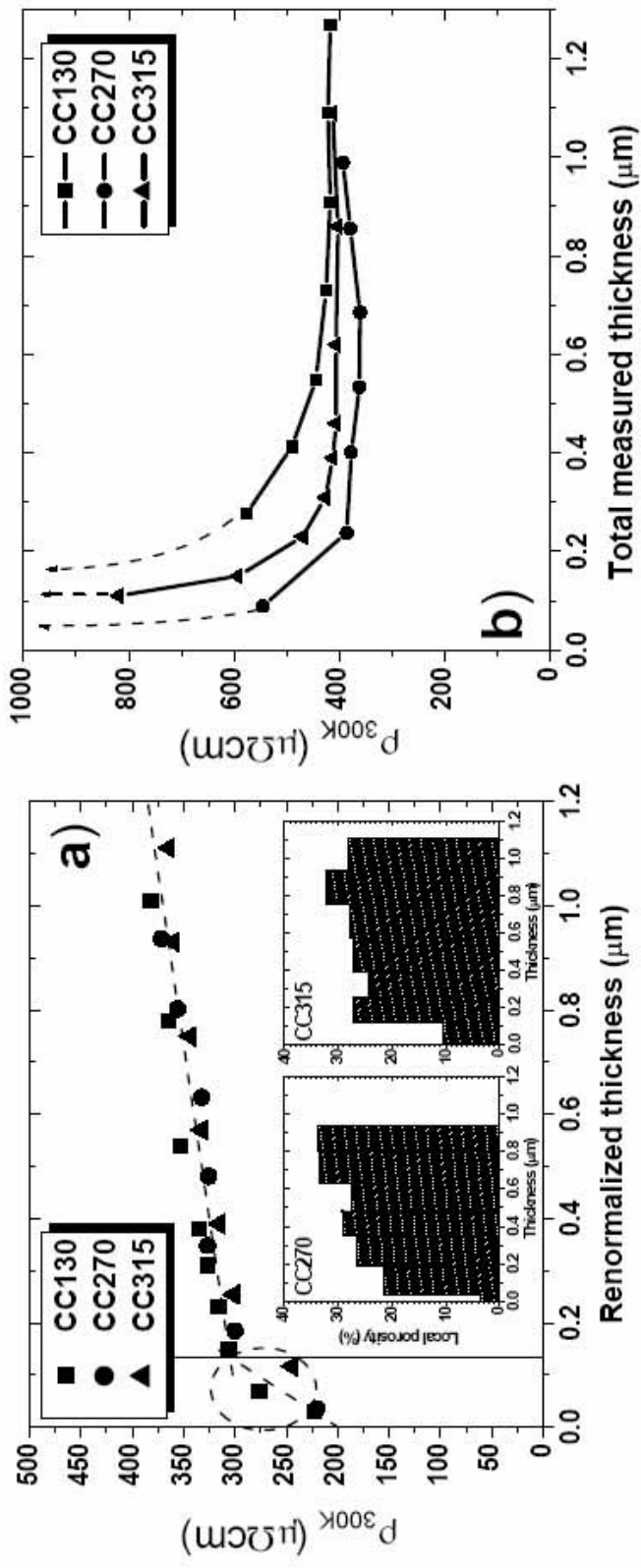

Fig 4. S.I. Kim et al.

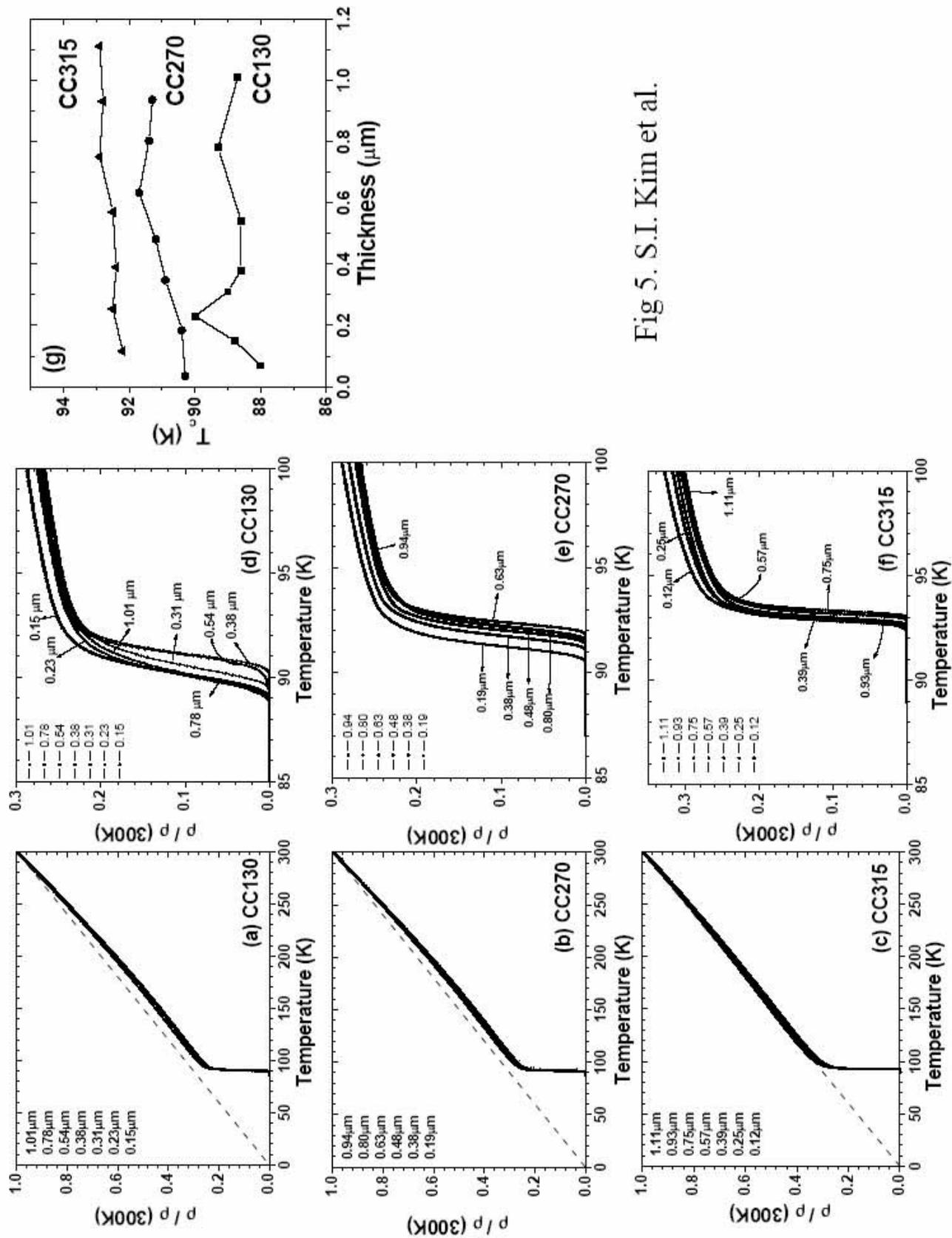

Fig 5. S.I. Kim et al.

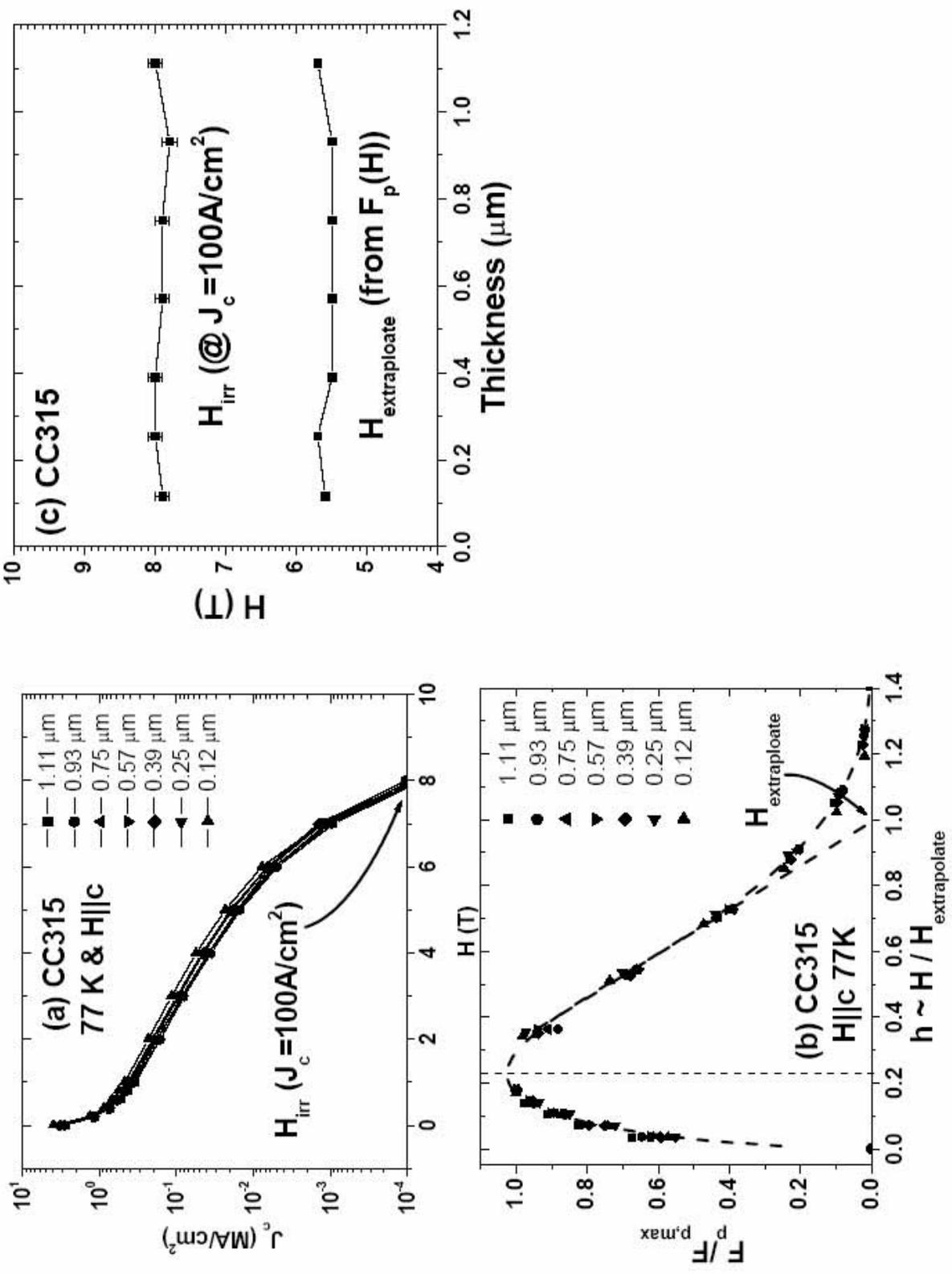

Fig 6. S.I. Kim et al.

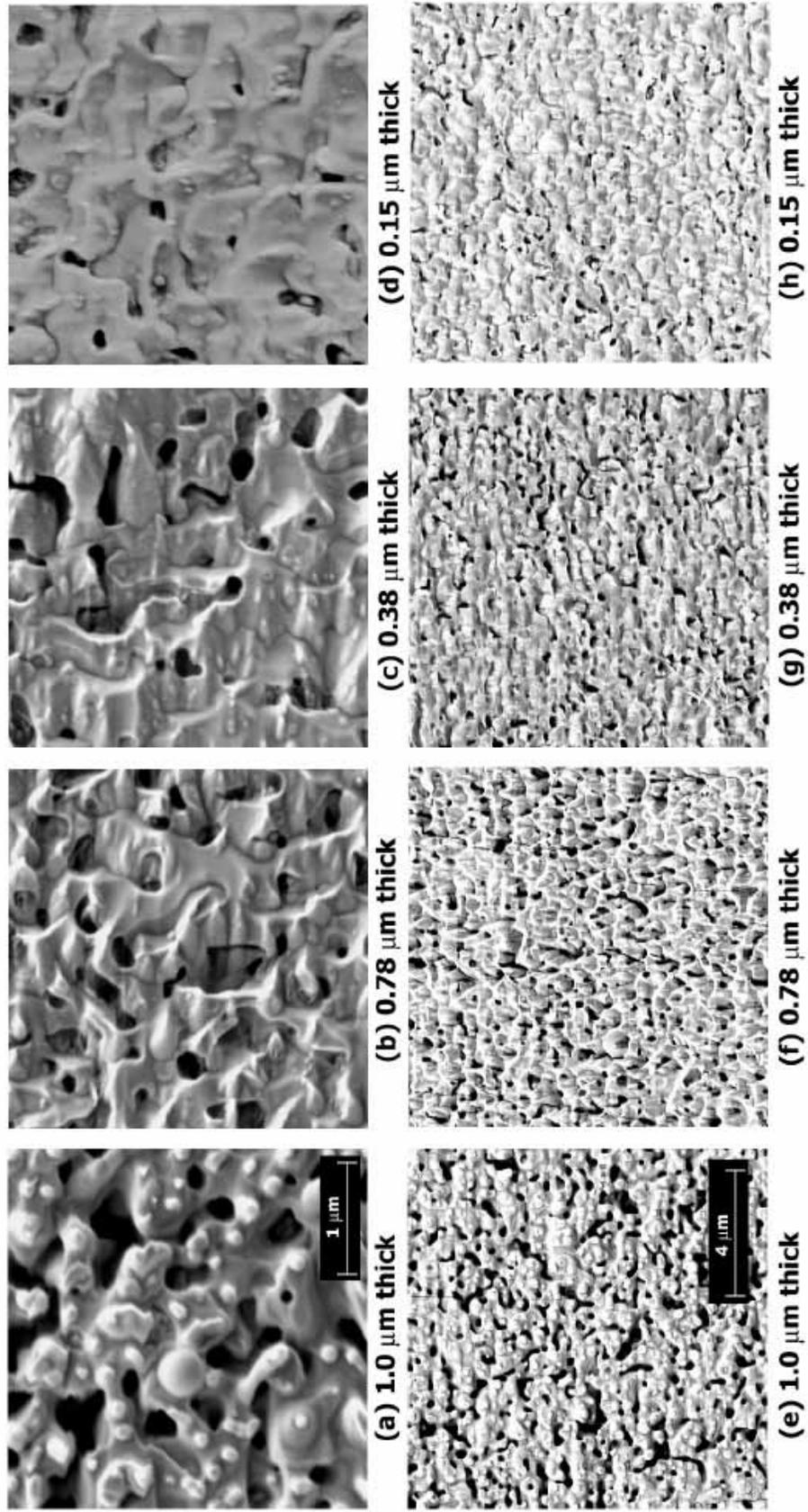

Fig 7. S.I. Kim et al.

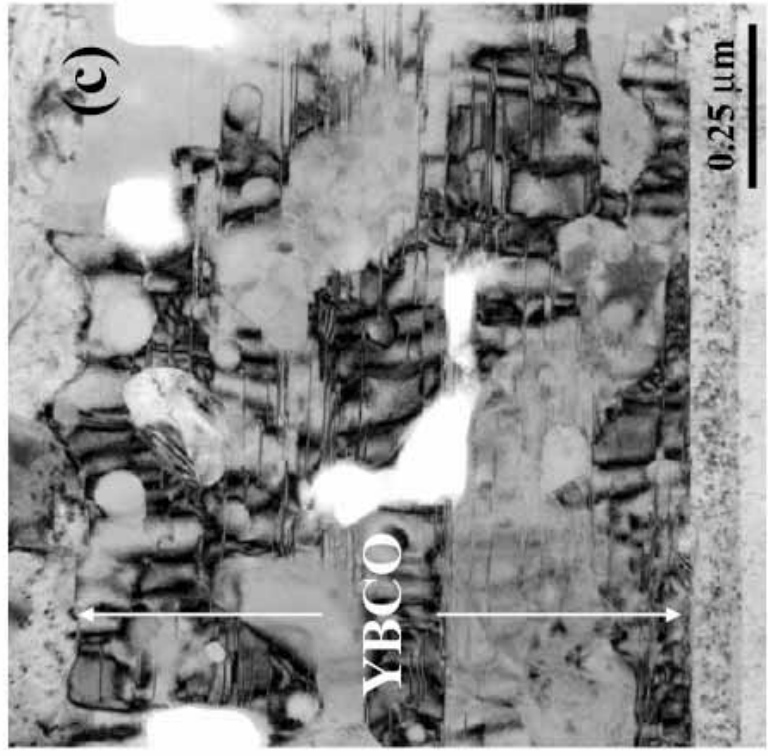
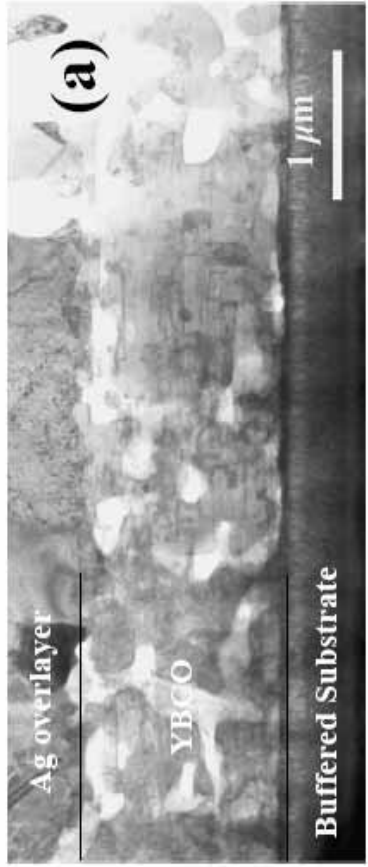
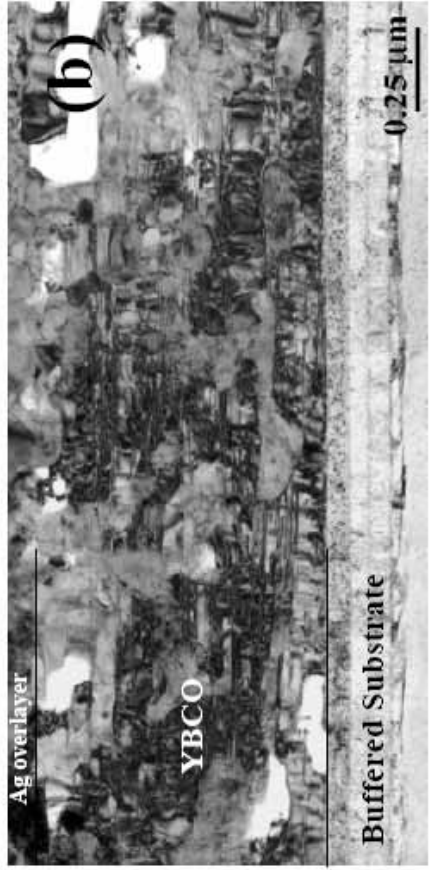

Fig 8. S.I. Kim et al.

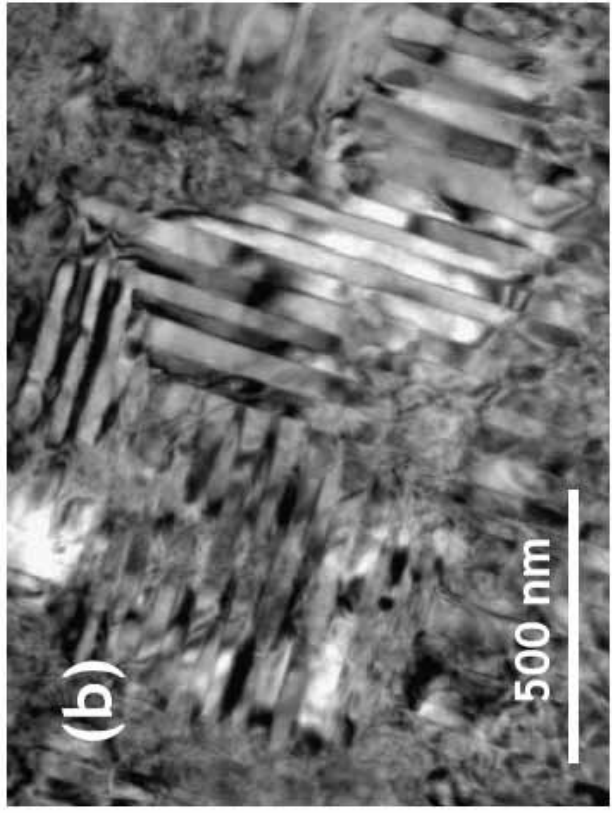

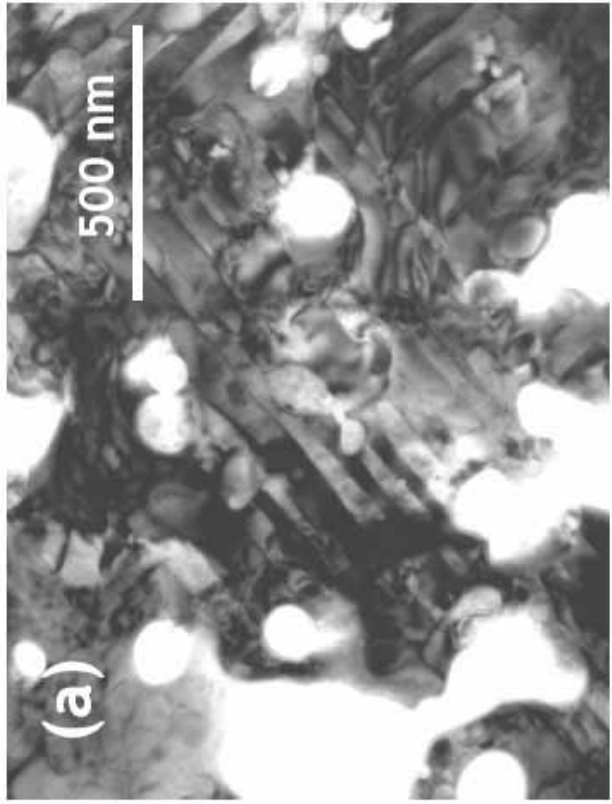

Fig 9. S.I. Kim et al.

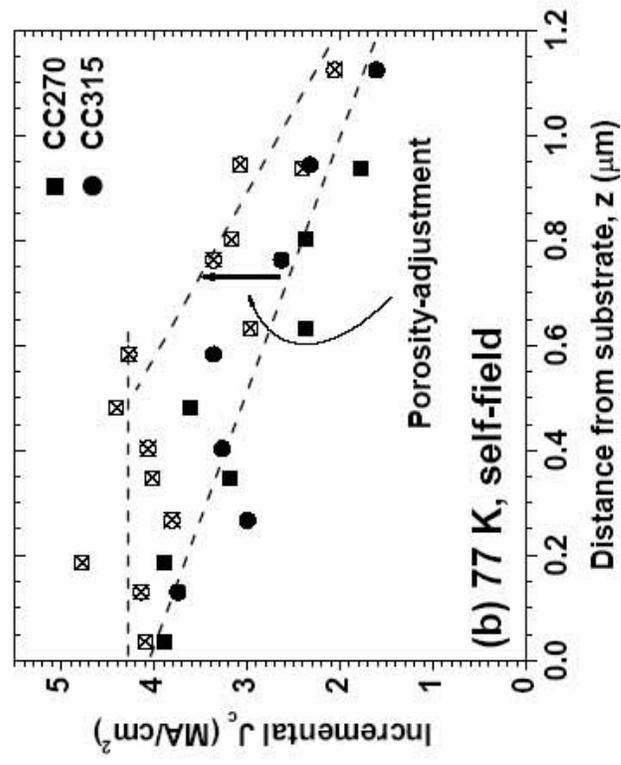
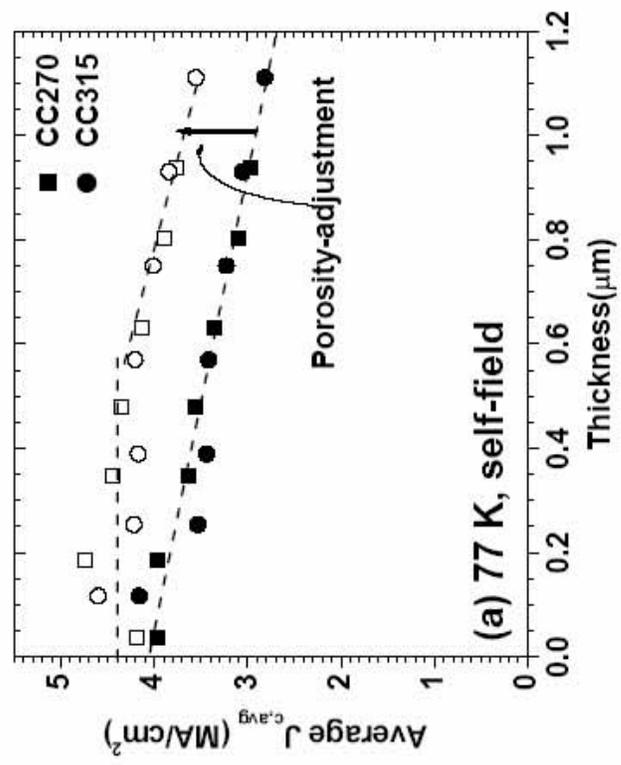

Fig 10. S.I. Kim et al.

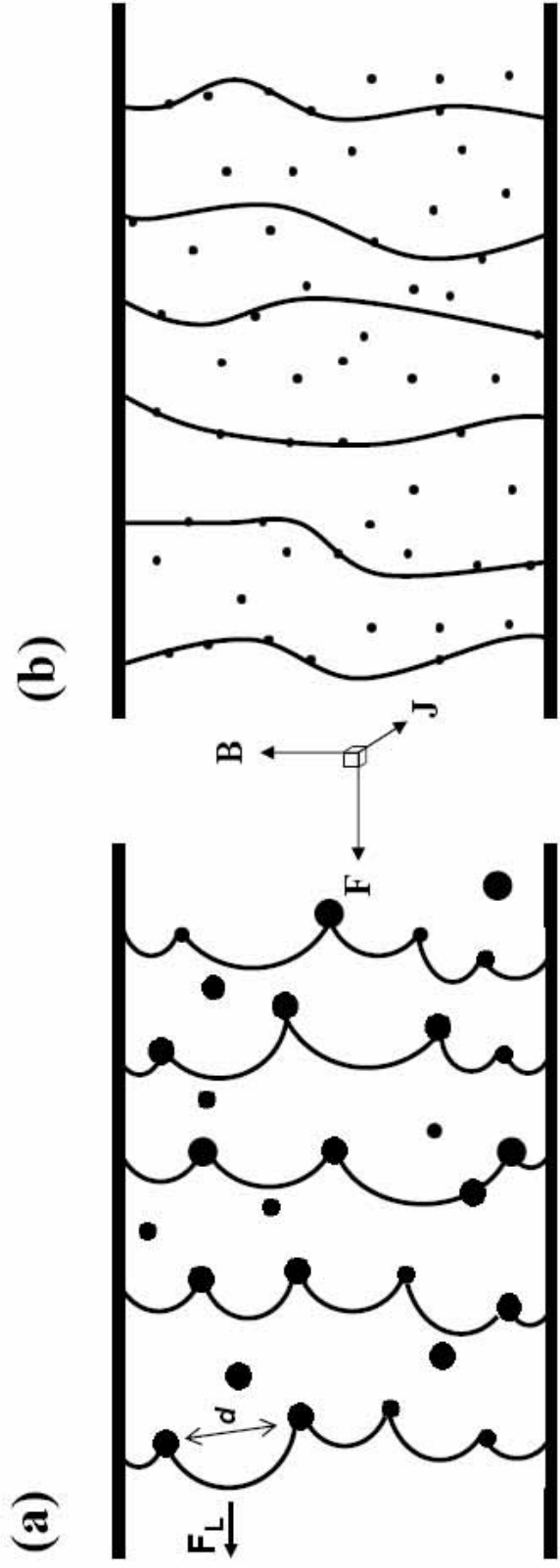

Fig 11. S.I. Kim et al.